 \newlength\smallfigwidth
\documentclass[aps,prb,twocolumn,floatfix,showpacs,amsmath,
amssymb ] {
revtex4-1}

\usepackage{amsmath}    
\usepackage{graphicx}   
\usepackage{verbatim}   
\usepackage{color}      
\usepackage{subfigure}  
\usepackage{float}
\usepackage{hyperref}   

\begin{document}


\title{Tuning magnetic charge population and mobility in unidirectional array of nanomagnets as a function of lattice parameters}
\author{R.\ S.\ Gon\c{c}alves}
\affiliation{ Laboratory of Spintronics and nanomagnetism (LabSpiN), Departamento de F\'{i}sica,
Universidade Federal de Vi\c cosa, Vi\c cosa, 36570-900, Minas Gerais, Brazil }
\author{R.\ P.\  Loreto}
\affiliation{ Laboratory of Spintronics and nanomagnetism (LabSpiN), Departamento de F\'{i}sica,
Universidade Federal de Vi\c cosa, Vi\c cosa, 36570-900, Minas Gerais, Brazil }
\author{J. Borme}
\affiliation{INL-International Iberian Nanotechnology Laboratory, 4715-330, Braga, Portugal}
\author{P.P. Freitas}
\affiliation{INL-International Iberian Nanotechnology Laboratory, 4715-330, Braga, Portugal}
\author{C.\ I.\ L.\ Araujo}
\email{dearaujo@ufv.br} \affiliation{ Laboratory of Spintronics and nanomagnetism (LabSpiN), Departamento de F\'{i}sica,
Universidade Federal de Vi\c cosa, Vi\c cosa, 36570-900, Minas Gerais, Brazil }
\date{\today}

\begin{abstract}
Sets of nanomagnets are often utilized to mimic cellular automata in design of nanomagnetic logic devices or frustration and emergence of magnetic charges in artificial spin ice systems. in previous work we showed that unidirectional arrangement of nanomagnets can behave as artificial spin ice, with frustration arising from second neighbor dipolar interaction, and present good magnetic charge mobility due to the low string tension among charges. Here, we present an experimental investigation of magnetic charge population and mobility in function of lateral and longitudinal distance among nanomagnets. Our results corroborate partially the theoretical predictions, performed elsewhere by emergent interaction model, could be useful in nanomagnet logic devices design and brings new insights about the best design for magnetic charge ballistic transport under low external magnetic field with magnetic charge mobility tunning for application in magnetricity.      
\end{abstract}

\maketitle

\section{introduction}

With the recent improvement in nanofabrication techniques, allowing development of samples composed by elements of few nanometers size, well organized and distributed in large areas, several new technologies based on nanodevices could be acquired, such as the huge improvement in computer processors scalability \cite{matzke1997will} or the utilization of metamaterials for nanophotonics development \cite{xiao2010loss}. In the field of magnetism, nanofabrication processes allowed the manipulation of magnetic domain structures, in general due to the combination of materials magnetic parameters with the shape anisotropy provided by the different nanostructures, enabling huge increase in information storage density and reading speed. Many examples can be highlighted, as the increase in capability of information storage density in magnetic nanodots \cite{shinjo2000magnetic} or antidots \cite{de2014magnetic, ribeiro2016investigation}, development of non-volatile magnetoresistive random access memories with in-plane \cite{khvalkovskiy2013basic} and out-of-plane \cite{perrissin2018highly} single domain magnetization, achieved in stacks of few nanometers diameter, multibits stored in single stack by magnetic vortex configuration \cite{de2016multilevel} and very stable and fast information transport allowed by the magnetic skyrmions created and manipulated in racetracks \cite{fert2013skyrmions, woo2016observation, loreto2017creation, zhang2016magnetic}. Other interesting features could be achieved by nanofabrication of magnetic nanobars with single domain magnetization, arranged precisely in several geometries in large areas, to replicate quantum cellular automata \cite{lent1993quantum} behavior in nanomagnetic logic devices \cite{atulasimha2010bennett, lambson2011exploring}, or realization and investigation of frustration in artificial spin ice systems \cite{wang2006artificial} at room temperature, due to stability of nanoisland magnetization to thermal fluctuations. Such a system was proposed to serve as feasible mimetization of pyrochlore natural spin ices \cite{ramirez1999zero, castelnovo2008magnetic}, which present at very low temperatures (below 1K) frustration of its magnetic moment. This material group name was given due to the similarity in frustration of its atomic magnetic moment with the atomic position in water ice. In the artificial spin ices, geometrical frustration generated artificially by the nanomagnets distribution, favors the ground state symmetry break by external excitations and emergency of magnetic charges in the vertex among nanomagnets, behaving like magnetic monopoles \cite{mol2009magnetic, ladak2010direct}. We have demonstrated in preliminary work \cite{loreto2015emergence} that a particular unidirectional arrangement of nanomagnets also behaves like artificial spin ice, due to frustrations with second neighbors, with generation of magnetic charge pairs and energetic string connecting them. In our investigations we showed that the string energy in the unidirectional spin ice system is low compared with the low energetic string observed in tridimensional artificial spin ice \cite{chern2014realizing, perrin2016extensive} and orthorhombic rectangular artificial spin ice \cite{nascimento2012confinement, ribeiro2017realization}. In this particular rectangular geometry, the dipolar interaction competition among horizontal and vertical nanomagnets in function of lattice stretch, brings the system to degeneracy in the particular orthorhombic configuration with higher probability of magnetic charge emergence and decrease in magnetic charge interaction, giving them more freedom to move under external magnetic field. The advantage of unidirectional artificial spin ice, in comparison with the tridimensional or orthorhombic, is the fact that the magnetic charge pairs can travel in a ballistic way through the magnetic nanowires formed by successive longitudinal nanomagnets, then its movements can be easily mapped and manipulated. In this paper we aim to experimentally investigate, through magnetic force microscopy measurements combined with external magnetic field application, the ground state magnetization and magnetic charge emergence probability, as well, the magnetic charge population and magnetization behavior under external magnetic field sweep. Such investigation is necessary for better understanding of unidirectional lattice influence in the system ground state, magnetic charge population and response to the external field for further application in magnetricity \cite{bramwell2012magnetic, blundell2012monopoles} devices or even for improvements in design of nanomagnetic logic devices.

\section{Experimental Methods}

For the nanofabrication of the investigated samples, thin film of ferromagnetic permalloy 20nm, preceded by tantalum 3nm for adhesion and followed by the same thickness of tantalum for capping, were growth by sputtering technique on silicon substrate. The different linear arrangements of nanomagnets with dimension of 3 $\mu$m x 400 nm, larger enough to have good magnetic signal and image contrast without losing its single domain magnetization, as carefully investigated with magnetic force microscopy by A. Imre and reported by W. Porod et al. \cite{porod2014nanomagnet}, were performed by electron beam lithography developed in 80nm of negative resist ARN7520 spin-coated over the substrate and backed in hot plate for 1 min under 85 $^{\circ}$C. Finally the samples were defined by ion milling etching of the unprotected thin film and subsequent oxygen plasma aching for the electroresist removal.

\begin{figure}[h!]
    \centering
   \includegraphics[width=0.49\textwidth]{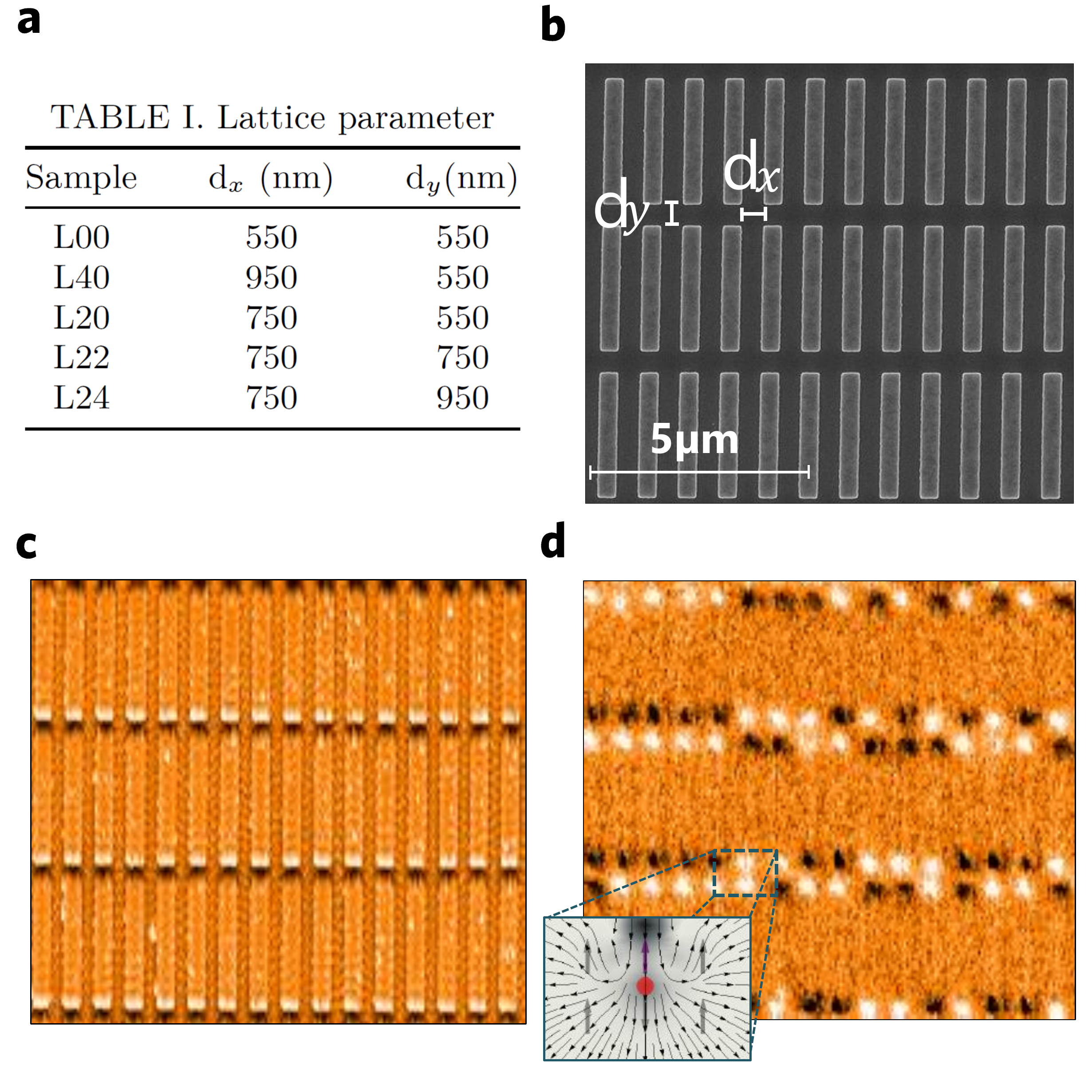}
    \caption{a) Table presenting the different samples investigated as function of nanomagnets lateral ($d_x$) and frontal ($d_y$) distances. b) Scanning electron image obtained in the nanofabricated sample showing good definition of the designed rectangular nanomagnets. c) atomic force microscopy topographic image of the sample and d) magnetic force image of same sample presenting the dipoles of the nanomagnets and the magnetic charges (bright and dark spots) emerged in the vertex. The inset show the simulated stray field generated by the nanobars and the magnetic charge field very similar to the one expected for point like magnetic monopole.}
    \label{fig:fig1}
\end{figure}

The different lattice parameters utilized for the unidirectional arrangements of nanomagnets are presented in the table of figure 1a, where $d_x$ is the lateral separation among nanomagnets while $d_y$ is the longitudinal separation. The good reproducibility of the nanomagnet rectangular design to the sample and good quality after milling process was characterized by field emission gun scanning electron microscopy, as in the image showed in Figure 1b. The topography (Figure 1c) and magnetization of the samples were characterized by magnetic force microscopy upgraded by an electromagnet for the external magnetic field application. The ground state of different samples, as in the example presented in Figure 1d, were obtained after demagnetization protocol performed by alternate magnetic field with amplitude of 1 kOe and frequency of 60 Hz applied in the y-direction, while the sample was slowly moved away from the coil center. For the hysteresis measurement each step of the field was applied in the y-direction before the magnetic force measurements.              

\section{Results and Discussions}

Without the second neighbors influence due to dipolar interaction, two nanomagnets would behave like macroscopic magnets, when aligned in longitudinal configuration they will have dipoles aligned in parallel (north to south), while when disposed laterally they will present antiparallel alignment of dipoles. Our main goal in this work is to investigate how changes in lateral and longitudinal distances among nanomagnets can affect the ground state of nanomagnets set behavior and also how this will affect the magnetic charge population and system response to magnetic field. First we have investigated the ground state in each of the five different lattices, from the magnetic force microscopy measurements, performed after demagnetization process. The analyses was made by counting the total number of magnetic charges and system magnetization from the dipoles direction.
The data obtained from the five different lattices are presented in Figure 2, with normalized magnetization in y-direction showed in Figure 2a and percentage of magnetic charge divided by the vertex number in Figure 2b. 
From the measurements summary in Figure 2a, it is possible to notice that the lowest magnetization $M_y$=0.064 was achieved in the lattice L20, the one with intermediate separation among nanomagnets in y-direction, increasing monotonically with the distance among nanomagnets in x-direction. Howbeit, the analysis of magnetic charge population in Figure 2b shows that in this L20 particular sample, together with the L40, the ground state is presenting the highest magnetic charge population, which implies that the magnetization decrease occurs likely with antiparallel aligment of longitudinal nanomagnets, condition to emergence of vertex magnetic charge, instead of lateral as expected. The sample presenting more occurrence of lateral antiferromagnetic alignment, with low magnetization combined with low magnetic charge population is the L22, but even so with reasonable number of ferromagnetic domains that carry magnetic charges emergence in its extremities.

\begin{figure}[h!]
    \centering
    \includegraphics[width=0.4\textwidth]{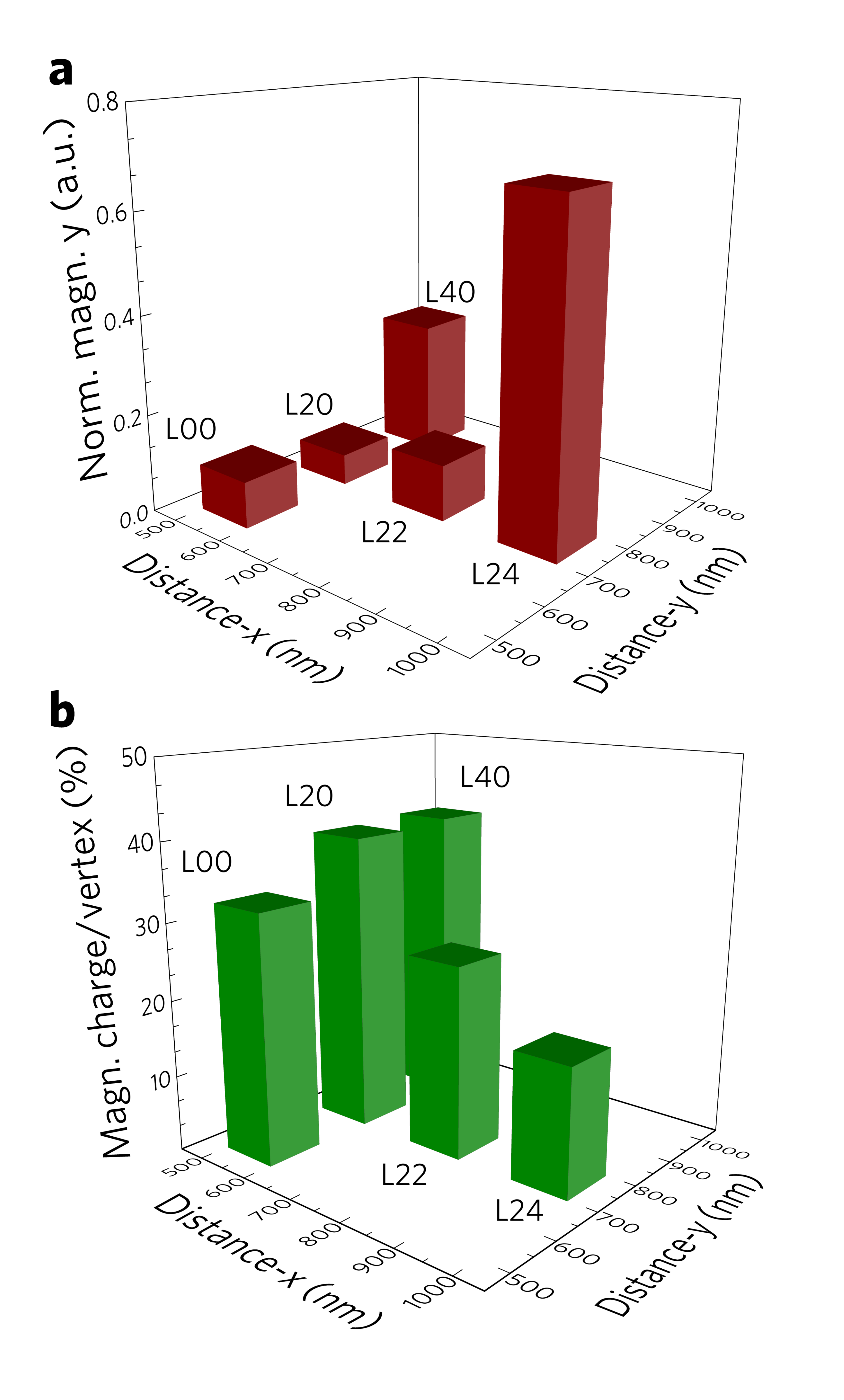}
    \caption{a) Normalized magnetization in y-direction and b) magnetic charge percentage divided per vertex as a function of frontal and lateral separations for all samples after demagnetization process.}
    \label{fig:fig2}
\end{figure}
  
The next step is to investigate the nanomagnets response to the external magnetic field in function of longitudinal and lateral separation, for that we have applied field in order to saturate the magnetization in y-direction and started to apply magnetic field to the opposite direction. Apart from the nanomagnets in the sample border, each nanomagnet magnetization flip will be responsible for a pair of magnetic charge creation. The flip of successive neighbor nanomagnets will separate the pair in a movement of opposite magnet charges that can be seen as a ballistic magnetic charge current. In a separate region of same nanomagnet wire composed by a set of longitudinal nanoislands, another nanomagnet magnetization flip can occur with creation of another magnetic charge pair. By increasing the magnetic field and promoting pair separations, opposite magnetic charges will meet and annihilate themselves. So, successive magnetic charge creation and annihilation will occurs until magnetization saturation to the opposite side. We aim to investigate what is the role of the lattice parameter in that process, giving more freedom to magnetic charges and decreasing the external magnetic field needed for its transport, in order to design low power consumption systems that can be inserted in the actual technology for magnetic field application in integrated circuits \cite{prinz1998magnetoelectronics}, allowing its further application in magnetricity or nanomagnetic  logic devices. The described process is depicted in Figure 3a, which is composed by magnetic force microscopy images taken in intermediate process of external magnetic field application and cartoon of magnetic charge mapped in the highlighted regions. The hysteresis curve with normalized magnetization in y-direction, obtained by nanomagnet dipole counting and half loop symmetric mirrored, is shown in Figure 3b, while the percentage of magnetic charges mapped divided per total vertex number is shown in Figure 3c.          
The summary of results obtained in the five different investigated lattices, following the sweep magnetic field protocol described above, are presented in Figure 4. The hysteresis obtained in the samples with unidirectional nanomagnet lattice stretched in the y-direction, same direction of external magnetic field application, is presented in Figure 4a, while the hysteresis of samples with separation in x-direction are presented in Figure 4b. From those results it is possible to notice that coercivity is totally affected by longitudinal separation among nanomagnets, increasing monotonically from $H_c=368 Oe$ to $H_c=464 Oe$. In the other configuration, we notice that the coercivity have been decreased from $H_c=426 Oe$, in the configuration with closer lateral distance among nanomagnets, to $H_c=390 Oe$ in the larger lateral distance configuration. From these results we could observe that lowest magnetic field will be needed to create and transport magnetic charges in lattices with lower longitudinal distances and higher lateral separation. As the amplitude of saturation field measured here is very close to the ones applied by bit lines in Toggle memories \cite{engel20054}, designs of a single wire composed by longitudinal nanomagnets can be used for ballistic transport with magnetic charge mobility being tunned by insertion of lateral wires.  

\begin{figure*}
	\includegraphics[width=0.75\textwidth]{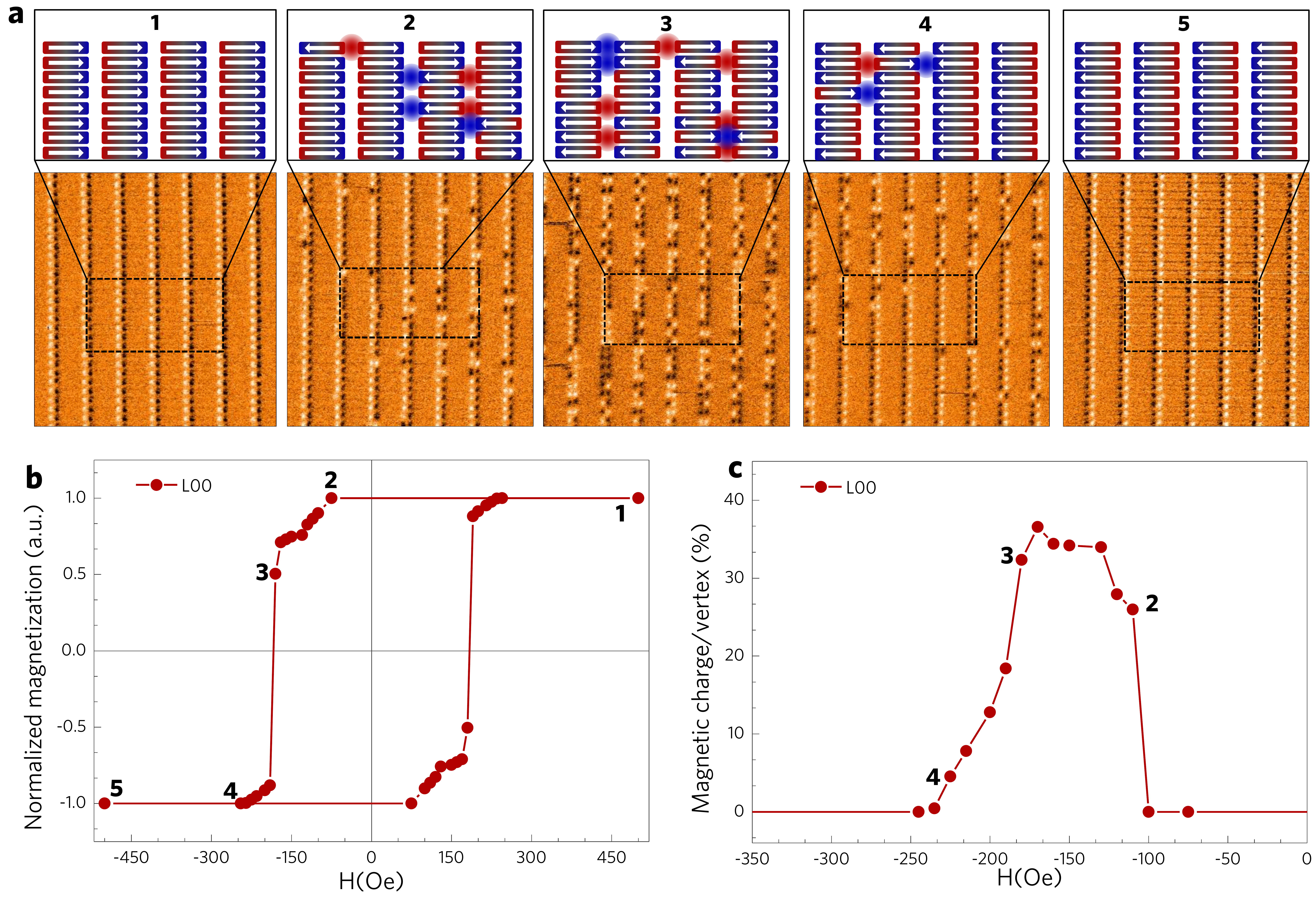}
	\caption{a) Magnetic force microscopy images of L00 sample, presenting magnetization of nanomagnets dipoles and magnetic charges in function of external magnetic field sweep. In the inset the evolution of magnetic charge emergence is represented by blue and red dots. b) Magnetic hysteresis loop obtained by counting nanomagnets magnetization. The numbers 1-5 represents the magnetic field applied before magnetic force measurements presented in figure a. c) Magnetic charge population as a function of external magnetic field applied extracted from Figure a.}
\end{figure*}
\begin{figure}
    \centering
   \includegraphics[width=0.48\textwidth]{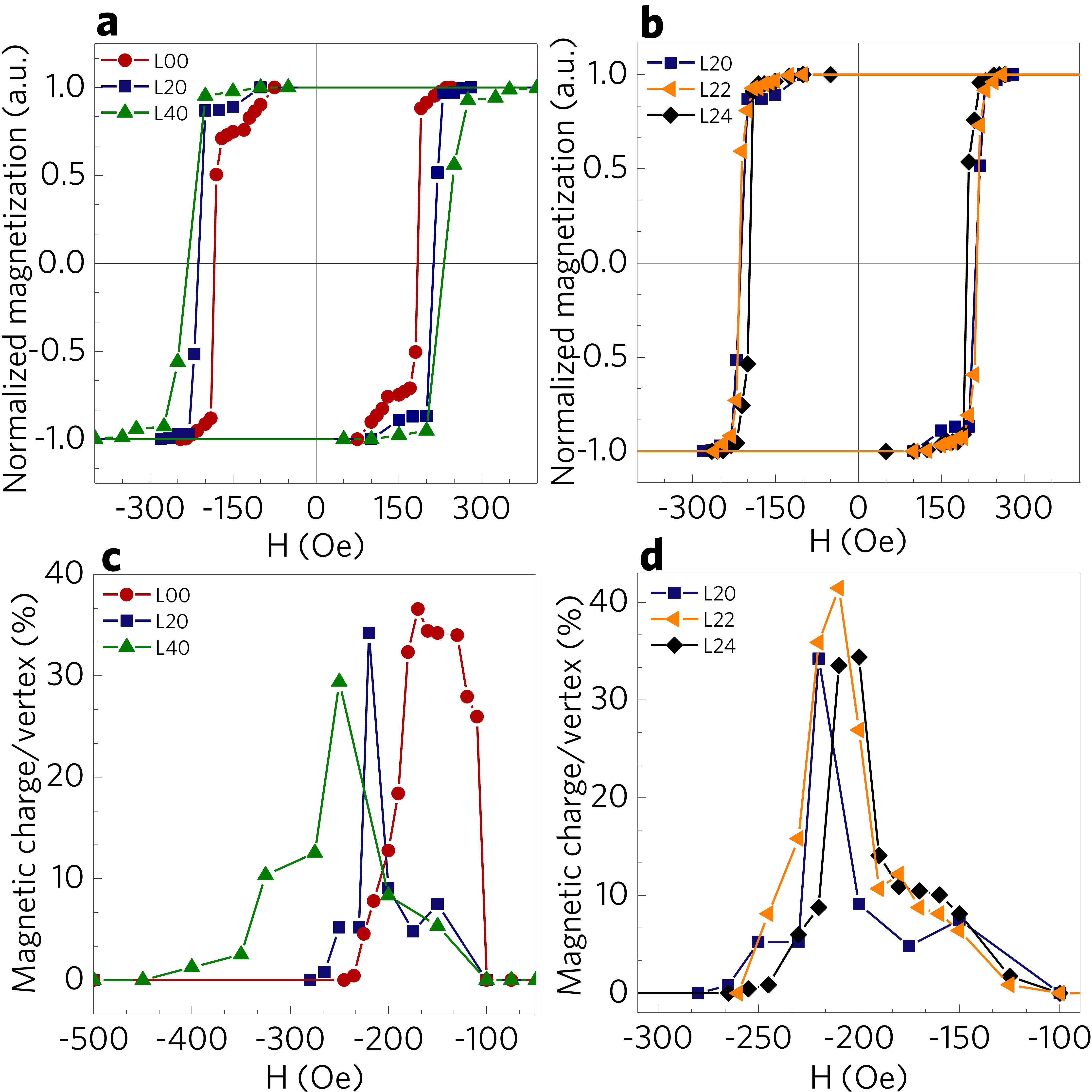}
    \caption{Magnetization hysteresis loop (M(H)) for the samples with a) different lateral separations ($ d_{x} $) and b) different frontal separations ($ d_{y} $) (see table I). c)  Magnetic charge percentage as a function of external applied field for the samples with c) different lateral separations and d) different frontal separations.}
    \label{fig:fig1}
\end{figure}
Finally, the evolution of magnetic charge population in function of external field and lattice parameters is presented in Figure 4c for the longitudinal separations and in Figure 4d for lateral separation among nanomagnets. These results show largest magnetic charge population at lower external fields, in lattice with lower separation in y-direction, and a decrease in magnetic charge population with peak dislocated to higher external magnetic fields in function of lattice stretch in y-direction.  We notice that our results corroborate in part the recent theoretical prediction by J.H. Rodrigues and L.A.S. M\'ol \cite{rodrigues2018towards}. In that work they have proposed a model based on interacting excitations to predict magnetic charge population in function of external field and magnetic charge intensity. They found out that the peak for magnetic charge population occurs at lower field for charges with higher intensity and is dislocate to higher fields for charges with lower intensity. That is similar to what we observe experimentally in the present work, considering that the magnetic charge emerging in lowest distances among longitudinal nanomagnets have more intensity than in larger distances among longitudinal nanomagnets, due to higher density of field lines. However, the monopoles population amount predicted by the simulations is behaving exactly in opposite way in comparison to the observed in this present work. We attribute such discrepancy to the fact that the theoretical analysis was based in constant string tension. In the present experimental observations not only the charge intensity is changing with lattice stretching, but also the string energy as observed in previous work in similar geometry \cite{loreto2015emergence}. That can be seen in Figure 4d where the magnetic charge intensity is the same for every lattice with same longitudinal magnets separation, but magnetic charge population is higher for mean lateral separation probably due to lowering in string tension by second neighbor influence. 

\section{Conclusion}   

Due to the second neighbors dipolar interaction, the antiferromagnetic ground state expected from first neighbors analysis was not verified, after desmagnetization protocol, in any nanomagnets array investigated in this work. Extrapolations of our results can point for such configuration with even lower lateral and longitudinal distance among nanomagnets. The lattice approaching more the antiferromagnetic configuration with lower number of magnetic charges was the L22, which also presented higher magnetic charge population with higher mobility under external magnetic field. So, we conclude that this geometry should present lower string tension among magnetic charges. The low field needed for nanomagnets magnetization saturation, presented in this work, is compatible for magnetricity device insertion in current circuit integration technology and the analysis of lattices with different lateral separations, suggest that the magnetic charge mobility can be tunned by changing lateral distances among lateral nanowires, in such kind of devices.   

\begin{acknowledgments}
The authors thank CNPq, CAPES and FAPEMIG (Brazilian agencies) for
financial support. 
\end{acknowledgments}

\bibliography{rafael}

\begin{thebibliography}{31}%
\makeatletter
\providecommand \@ifxundefined [1]{%
 \@ifx{#1\undefined}
}%
\providecommand \@ifnum [1]{%
 \ifnum #1\expandafter \@firstoftwo
 \else \expandafter \@secondoftwo
 \fi
}%
\providecommand \@ifx [1]{%
 \ifx #1\expandafter \@firstoftwo
 \else \expandafter \@secondoftwo
 \fi
}%
\providecommand \natexlab [1]{#1}%
\providecommand \enquote  [1]{``#1''}%
\providecommand \bibnamefont  [1]{#1}%
\providecommand \bibfnamefont [1]{#1}%
\providecommand \citenamefont [1]{#1}%
\providecommand \href@noop [0]{\@secondoftwo}%
\providecommand \href [0]{\begingroup \@sanitize@url \@href}%
\providecommand \@href[1]{\@@startlink{#1}\@@href}%
\providecommand \@@href[1]{\endgroup#1\@@endlink}%
\providecommand \@sanitize@url [0]{\catcode `\\12\catcode `\$12\catcode
  `\&12\catcode `\#12\catcode `\^12\catcode `\_12\catcode `\%12\relax}%
\providecommand \@@startlink[1]{}%
\providecommand \@@endlink[0]{}%
\providecommand \url  [0]{\begingroup\@sanitize@url \@url }%
\providecommand \@url [1]{\endgroup\@href {#1}{\urlprefix }}%
\providecommand \urlprefix  [0]{URL }%
\providecommand \Eprint [0]{\href }%
\providecommand \doibase [0]{http://dx.doi.org/}%
\providecommand \selectlanguage [0]{\@gobble}%
\providecommand \bibinfo  [0]{\@secondoftwo}%
\providecommand \bibfield  [0]{\@secondoftwo}%
\providecommand \translation [1]{[#1]}%
\providecommand \BibitemOpen [0]{}%
\providecommand \bibitemStop [0]{}%
\providecommand \bibitemNoStop [0]{.\EOS\space}%
\providecommand \EOS [0]{\spacefactor3000\relax}%
\providecommand \BibitemShut  [1]{\csname bibitem#1\endcsname}%
\let\auto@bib@innerbib\@empty
\bibitem [{\citenamefont {Matzke}(1997)}]{matzke1997will}%
  \BibitemOpen
  \bibfield  {author} {\bibinfo {author} {\bibfnamefont {D.}~\bibnamefont
  {Matzke}},\ }\href@noop {} {\bibfield  {journal} {\bibinfo  {journal}
  {Computer}\ }\textbf {\bibinfo {volume} {30}},\ \bibinfo {pages} {37}
  (\bibinfo {year} {1997})}\BibitemShut {NoStop}%
\bibitem [{\citenamefont {Xiao}\ \emph {et~al.}(2010)\citenamefont {Xiao},
  \citenamefont {Drachev}, \citenamefont {Kildishev}, \citenamefont {Ni},
  \citenamefont {Chettiar}, \citenamefont {Yuan},\ and\ \citenamefont
  {Shalaev}}]{xiao2010loss}%
  \BibitemOpen
  \bibfield  {author} {\bibinfo {author} {\bibfnamefont {S.}~\bibnamefont
  {Xiao}}, \bibinfo {author} {\bibfnamefont {V.~P.}\ \bibnamefont {Drachev}},
  \bibinfo {author} {\bibfnamefont {A.~V.}\ \bibnamefont {Kildishev}}, \bibinfo
  {author} {\bibfnamefont {X.}~\bibnamefont {Ni}}, \bibinfo {author}
  {\bibfnamefont {U.~K.}\ \bibnamefont {Chettiar}}, \bibinfo {author}
  {\bibfnamefont {H.-K.}\ \bibnamefont {Yuan}}, \ and\ \bibinfo {author}
  {\bibfnamefont {V.~M.}\ \bibnamefont {Shalaev}},\ }\href@noop {} {\bibfield
  {journal} {\bibinfo  {journal} {Nature}\ }\textbf {\bibinfo {volume} {466}},\
  \bibinfo {pages} {735} (\bibinfo {year} {2010})}\BibitemShut {NoStop}%
\bibitem [{\citenamefont {Shinjo}\ \emph {et~al.}(2000)\citenamefont {Shinjo},
  \citenamefont {Okuno}, \citenamefont {Hassdorf}, \citenamefont {Shigeto},\
  and\ \citenamefont {Ono}}]{shinjo2000magnetic}%
  \BibitemOpen
  \bibfield  {author} {\bibinfo {author} {\bibfnamefont {T.}~\bibnamefont
  {Shinjo}}, \bibinfo {author} {\bibfnamefont {T.}~\bibnamefont {Okuno}},
  \bibinfo {author} {\bibfnamefont {R.}~\bibnamefont {Hassdorf}}, \bibinfo
  {author} {\bibfnamefont {K.}~\bibnamefont {Shigeto}}, \ and\ \bibinfo
  {author} {\bibfnamefont {T.}~\bibnamefont {Ono}},\ }\href@noop {} {\bibfield
  {journal} {\bibinfo  {journal} {Science}\ }\textbf {\bibinfo {volume}
  {289}},\ \bibinfo {pages} {930} (\bibinfo {year} {2000})}\BibitemShut
  {NoStop}%
\bibitem [{\citenamefont {de~Araujo}\ \emph {et~al.}(2014)\citenamefont
  {de~Araujo}, \citenamefont {Silva}, \citenamefont {Ribeiro}, \citenamefont
  {Nascimento}, \citenamefont {Felix}, \citenamefont {Ferreira}, \citenamefont
  {M{\'o}l}, \citenamefont {Moura-Melo},\ and\ \citenamefont
  {Pereira}}]{de2014magnetic}%
  \BibitemOpen
  \bibfield  {author} {\bibinfo {author} {\bibfnamefont {C.}~\bibnamefont
  {de~Araujo}}, \bibinfo {author} {\bibfnamefont {R.}~\bibnamefont {Silva}},
  \bibinfo {author} {\bibfnamefont {I.}~\bibnamefont {Ribeiro}}, \bibinfo
  {author} {\bibfnamefont {F.}~\bibnamefont {Nascimento}}, \bibinfo {author}
  {\bibfnamefont {J.}~\bibnamefont {Felix}}, \bibinfo {author} {\bibfnamefont
  {S.}~\bibnamefont {Ferreira}}, \bibinfo {author} {\bibfnamefont
  {L.}~\bibnamefont {M{\'o}l}}, \bibinfo {author} {\bibfnamefont
  {W.}~\bibnamefont {Moura-Melo}}, \ and\ \bibinfo {author} {\bibfnamefont
  {A.}~\bibnamefont {Pereira}},\ }\href@noop {} {\bibfield  {journal} {\bibinfo
   {journal} {Applied Physics Letters}\ }\textbf {\bibinfo {volume} {104}},\
  \bibinfo {pages} {092402} (\bibinfo {year} {2014})}\BibitemShut {NoStop}%
\bibitem [{\citenamefont {Ribeiro}\ \emph {et~al.}(2016)\citenamefont
  {Ribeiro}, \citenamefont {Felix}, \citenamefont {Figueiredo}, \citenamefont
  {Morais}, \citenamefont {Ferreira}, \citenamefont {Moura-Melo}, \citenamefont
  {Pereira}, \citenamefont {Quindeau},\ and\ \citenamefont
  {de~Araujo}}]{ribeiro2016investigation}%
  \BibitemOpen
  \bibfield  {author} {\bibinfo {author} {\bibfnamefont {I.}~\bibnamefont
  {Ribeiro}}, \bibinfo {author} {\bibfnamefont {J.}~\bibnamefont {Felix}},
  \bibinfo {author} {\bibfnamefont {L.}~\bibnamefont {Figueiredo}}, \bibinfo
  {author} {\bibfnamefont {P.}~\bibnamefont {Morais}}, \bibinfo {author}
  {\bibfnamefont {S.}~\bibnamefont {Ferreira}}, \bibinfo {author}
  {\bibfnamefont {W.}~\bibnamefont {Moura-Melo}}, \bibinfo {author}
  {\bibfnamefont {A.}~\bibnamefont {Pereira}}, \bibinfo {author} {\bibfnamefont
  {A.}~\bibnamefont {Quindeau}}, \ and\ \bibinfo {author} {\bibfnamefont
  {C.}~\bibnamefont {de~Araujo}},\ }\href@noop {} {\bibfield  {journal}
  {\bibinfo  {journal} {Journal of Physics: Condensed Matter}\ }\textbf
  {\bibinfo {volume} {28}},\ \bibinfo {pages} {456002} (\bibinfo {year}
  {2016})}\BibitemShut {NoStop}%
\bibitem [{\citenamefont {Khvalkovskiy}\ \emph {et~al.}(2013)\citenamefont
  {Khvalkovskiy}, \citenamefont {Apalkov}, \citenamefont {Watts}, \citenamefont
  {Chepulskii}, \citenamefont {Beach}, \citenamefont {Ong}, \citenamefont
  {Tang}, \citenamefont {Driskill-Smith}, \citenamefont {Butler}, \citenamefont
  {Visscher} \emph {et~al.}}]{khvalkovskiy2013basic}%
  \BibitemOpen
  \bibfield  {author} {\bibinfo {author} {\bibfnamefont {A.}~\bibnamefont
  {Khvalkovskiy}}, \bibinfo {author} {\bibfnamefont {D.}~\bibnamefont
  {Apalkov}}, \bibinfo {author} {\bibfnamefont {S.}~\bibnamefont {Watts}},
  \bibinfo {author} {\bibfnamefont {R.}~\bibnamefont {Chepulskii}}, \bibinfo
  {author} {\bibfnamefont {R.}~\bibnamefont {Beach}}, \bibinfo {author}
  {\bibfnamefont {A.}~\bibnamefont {Ong}}, \bibinfo {author} {\bibfnamefont
  {X.}~\bibnamefont {Tang}}, \bibinfo {author} {\bibfnamefont {A.}~\bibnamefont
  {Driskill-Smith}}, \bibinfo {author} {\bibfnamefont {W.}~\bibnamefont
  {Butler}}, \bibinfo {author} {\bibfnamefont {P.}~\bibnamefont {Visscher}},
  \emph {et~al.},\ }\href@noop {} {\bibfield  {journal} {\bibinfo  {journal}
  {Journal of Physics D: Applied Physics}\ }\textbf {\bibinfo {volume} {46}},\
  \bibinfo {pages} {074001} (\bibinfo {year} {2013})}\BibitemShut {NoStop}%
\bibitem [{\citenamefont {Perrissin}\ \emph {et~al.}(2018)\citenamefont
  {Perrissin}, \citenamefont {Lequeux}, \citenamefont {Strelkov}, \citenamefont
  {Chavent}, \citenamefont {Vila}, \citenamefont {Buda-Prejbeanu},
  \citenamefont {Auffret}, \citenamefont {Sousa}, \citenamefont {Prejbeanu},\
  and\ \citenamefont {Dieny}}]{perrissin2018highly}%
  \BibitemOpen
  \bibfield  {author} {\bibinfo {author} {\bibfnamefont {N.}~\bibnamefont
  {Perrissin}}, \bibinfo {author} {\bibfnamefont {S.}~\bibnamefont {Lequeux}},
  \bibinfo {author} {\bibfnamefont {N.}~\bibnamefont {Strelkov}}, \bibinfo
  {author} {\bibfnamefont {A.}~\bibnamefont {Chavent}}, \bibinfo {author}
  {\bibfnamefont {L.}~\bibnamefont {Vila}}, \bibinfo {author} {\bibfnamefont
  {L.~D.}\ \bibnamefont {Buda-Prejbeanu}}, \bibinfo {author} {\bibfnamefont
  {S.}~\bibnamefont {Auffret}}, \bibinfo {author} {\bibfnamefont {R.~C.}\
  \bibnamefont {Sousa}}, \bibinfo {author} {\bibfnamefont {I.~L.}\ \bibnamefont
  {Prejbeanu}}, \ and\ \bibinfo {author} {\bibfnamefont {B.}~\bibnamefont
  {Dieny}},\ }\href@noop {} {\bibfield  {journal} {\bibinfo  {journal}
  {Nanoscale}\ }\textbf {\bibinfo {volume} {10}},\ \bibinfo {pages} {12187}
  (\bibinfo {year} {2018})}\BibitemShut {NoStop}%
\bibitem [{\citenamefont {De~Araujo}\ \emph {et~al.}(2016)\citenamefont
  {De~Araujo}, \citenamefont {Alves}, \citenamefont {Buda-Prejbeanu},\ and\
  \citenamefont {Dieny}}]{de2016multilevel}%
  \BibitemOpen
  \bibfield  {author} {\bibinfo {author} {\bibfnamefont {C.~I.~L.}\
  \bibnamefont {De~Araujo}}, \bibinfo {author} {\bibfnamefont {S.~G.}\
  \bibnamefont {Alves}}, \bibinfo {author} {\bibfnamefont {L.~D.}\ \bibnamefont
  {Buda-Prejbeanu}}, \ and\ \bibinfo {author} {\bibfnamefont {B.}~\bibnamefont
  {Dieny}},\ }\href@noop {} {\bibfield  {journal} {\bibinfo  {journal} {Phys.
  Rev. Appl.}\ }\textbf {\bibinfo {volume} {6}},\ \bibinfo {pages} {024015}
  (\bibinfo {year} {2016})}\BibitemShut {NoStop}%
\bibitem [{\citenamefont {Fert}\ \emph {et~al.}(2013)\citenamefont {Fert},
  \citenamefont {Cros},\ and\ \citenamefont {Sampaio}}]{fert2013skyrmions}%
  \BibitemOpen
  \bibfield  {author} {\bibinfo {author} {\bibfnamefont {A.}~\bibnamefont
  {Fert}}, \bibinfo {author} {\bibfnamefont {V.}~\bibnamefont {Cros}}, \ and\
  \bibinfo {author} {\bibfnamefont {J.}~\bibnamefont {Sampaio}},\ }\href@noop
  {} {\bibfield  {journal} {\bibinfo  {journal} {Nature nanotechnology}\
  }\textbf {\bibinfo {volume} {8}},\ \bibinfo {pages} {152} (\bibinfo {year}
  {2013})}\BibitemShut {NoStop}%
\bibitem [{\citenamefont {Woo}\ \emph {et~al.}(2016)\citenamefont {Woo},
  \citenamefont {Litzius}, \citenamefont {Kr{\"u}ger}, \citenamefont {Im},
  \citenamefont {Caretta}, \citenamefont {Richter}, \citenamefont {Mann},
  \citenamefont {Krone}, \citenamefont {Reeve}, \citenamefont {Weigand} \emph
  {et~al.}}]{woo2016observation}%
  \BibitemOpen
  \bibfield  {author} {\bibinfo {author} {\bibfnamefont {S.}~\bibnamefont
  {Woo}}, \bibinfo {author} {\bibfnamefont {K.}~\bibnamefont {Litzius}},
  \bibinfo {author} {\bibfnamefont {B.}~\bibnamefont {Kr{\"u}ger}}, \bibinfo
  {author} {\bibfnamefont {M.-Y.}\ \bibnamefont {Im}}, \bibinfo {author}
  {\bibfnamefont {L.}~\bibnamefont {Caretta}}, \bibinfo {author} {\bibfnamefont
  {K.}~\bibnamefont {Richter}}, \bibinfo {author} {\bibfnamefont
  {M.}~\bibnamefont {Mann}}, \bibinfo {author} {\bibfnamefont {A.}~\bibnamefont
  {Krone}}, \bibinfo {author} {\bibfnamefont {R.~M.}\ \bibnamefont {Reeve}},
  \bibinfo {author} {\bibfnamefont {M.}~\bibnamefont {Weigand}},  \emph
  {et~al.},\ }\href@noop {} {\bibfield  {journal} {\bibinfo  {journal} {Nature
  materials}\ }\textbf {\bibinfo {volume} {15}},\ \bibinfo {pages} {501}
  (\bibinfo {year} {2016})}\BibitemShut {NoStop}%
\bibitem [{\citenamefont {Loreto}\ \emph {et~al.}(2018)\citenamefont {Loreto},
  \citenamefont {Moura-Melo}, \citenamefont {Pereira}, \citenamefont {Zhang},
  \citenamefont {Zhou}, \citenamefont {Ezawa},\ and\ \citenamefont
  {de~Araujo}}]{loreto2017creation}%
  \BibitemOpen
  \bibfield  {author} {\bibinfo {author} {\bibfnamefont {R.~P.}\ \bibnamefont
  {Loreto}}, \bibinfo {author} {\bibfnamefont {W.~A.}\ \bibnamefont
  {Moura-Melo}}, \bibinfo {author} {\bibfnamefont {A.~R.}\ \bibnamefont
  {Pereira}}, \bibinfo {author} {\bibfnamefont {X.}~\bibnamefont {Zhang}},
  \bibinfo {author} {\bibfnamefont {Y.}~\bibnamefont {Zhou}}, \bibinfo {author}
  {\bibfnamefont {M.}~\bibnamefont {Ezawa}}, \ and\ \bibinfo {author}
  {\bibfnamefont {C.~I.}\ \bibnamefont {de~Araujo}},\ }\href@noop {} {\bibfield
   {journal} {\bibinfo  {journal} {Journal of Magnetism and Magnetic
  Materials}\ }\textbf {\bibinfo {volume} {455}},\ \bibinfo {pages} {25}
  (\bibinfo {year} {2018})}\BibitemShut {NoStop}%
\bibitem [{\citenamefont {Zhang}\ \emph {et~al.}(2016)\citenamefont {Zhang},
  \citenamefont {Zhou},\ and\ \citenamefont {Ezawa}}]{zhang2016magnetic}%
  \BibitemOpen
  \bibfield  {author} {\bibinfo {author} {\bibfnamefont {X.}~\bibnamefont
  {Zhang}}, \bibinfo {author} {\bibfnamefont {Y.}~\bibnamefont {Zhou}}, \ and\
  \bibinfo {author} {\bibfnamefont {M.}~\bibnamefont {Ezawa}},\ }\href@noop {}
  {\bibfield  {journal} {\bibinfo  {journal} {Nature communications}\ }\textbf
  {\bibinfo {volume} {7}},\ \bibinfo {pages} {10293} (\bibinfo {year}
  {2016})}\BibitemShut {NoStop}%
\bibitem [{\citenamefont {Lent}\ \emph {et~al.}(1993)\citenamefont {Lent},
  \citenamefont {Tougaw}, \citenamefont {Porod},\ and\ \citenamefont
  {Bernstein}}]{lent1993quantum}%
  \BibitemOpen
  \bibfield  {author} {\bibinfo {author} {\bibfnamefont {C.~S.}\ \bibnamefont
  {Lent}}, \bibinfo {author} {\bibfnamefont {P.~D.}\ \bibnamefont {Tougaw}},
  \bibinfo {author} {\bibfnamefont {W.}~\bibnamefont {Porod}}, \ and\ \bibinfo
  {author} {\bibfnamefont {G.~H.}\ \bibnamefont {Bernstein}},\ }\href@noop {}
  {\bibfield  {journal} {\bibinfo  {journal} {Nanotechnology}\ }\textbf
  {\bibinfo {volume} {4}},\ \bibinfo {pages} {49} (\bibinfo {year}
  {1993})}\BibitemShut {NoStop}%
\bibitem [{\citenamefont {Atulasimha}\ and\ \citenamefont
  {Bandyopadhyay}(2010)}]{atulasimha2010bennett}%
  \BibitemOpen
  \bibfield  {author} {\bibinfo {author} {\bibfnamefont {J.}~\bibnamefont
  {Atulasimha}}\ and\ \bibinfo {author} {\bibfnamefont {S.}~\bibnamefont
  {Bandyopadhyay}},\ }\href@noop {} {\bibfield  {journal} {\bibinfo  {journal}
  {Applied Physics Letters}\ }\textbf {\bibinfo {volume} {97}},\ \bibinfo
  {pages} {173105} (\bibinfo {year} {2010})}\BibitemShut {NoStop}%
\bibitem [{\citenamefont {Lambson}\ \emph {et~al.}(2011)\citenamefont
  {Lambson}, \citenamefont {Carlton},\ and\ \citenamefont
  {Bokor}}]{lambson2011exploring}%
  \BibitemOpen
  \bibfield  {author} {\bibinfo {author} {\bibfnamefont {B.}~\bibnamefont
  {Lambson}}, \bibinfo {author} {\bibfnamefont {D.}~\bibnamefont {Carlton}}, \
  and\ \bibinfo {author} {\bibfnamefont {J.}~\bibnamefont {Bokor}},\
  }\href@noop {} {\bibfield  {journal} {\bibinfo  {journal} {Physical review
  letters}\ }\textbf {\bibinfo {volume} {107}},\ \bibinfo {pages} {010604}
  (\bibinfo {year} {2011})}\BibitemShut {NoStop}%
\bibitem [{\citenamefont {Wang}\ \emph {et~al.}(2006)\citenamefont {Wang},
  \citenamefont {Nisoli}, \citenamefont {Freitas}, \citenamefont {Li},
  \citenamefont {McConville}, \citenamefont {Cooley}, \citenamefont {Lund},
  \citenamefont {Samarth}, \citenamefont {Leighton}, \citenamefont {Crespi}
  \emph {et~al.}}]{wang2006artificial}%
  \BibitemOpen
  \bibfield  {author} {\bibinfo {author} {\bibfnamefont {.~R.}\ \bibnamefont
  {Wang}}, \bibinfo {author} {\bibfnamefont {C.}~\bibnamefont {Nisoli}},
  \bibinfo {author} {\bibfnamefont {R.}~\bibnamefont {Freitas}}, \bibinfo
  {author} {\bibfnamefont {J.}~\bibnamefont {Li}}, \bibinfo {author}
  {\bibfnamefont {W.}~\bibnamefont {McConville}}, \bibinfo {author}
  {\bibfnamefont {B.}~\bibnamefont {Cooley}}, \bibinfo {author} {\bibfnamefont
  {M.}~\bibnamefont {Lund}}, \bibinfo {author} {\bibfnamefont {N.}~\bibnamefont
  {Samarth}}, \bibinfo {author} {\bibfnamefont {C.}~\bibnamefont {Leighton}},
  \bibinfo {author} {\bibfnamefont {V.}~\bibnamefont {Crespi}},  \emph
  {et~al.},\ }\href@noop {} {\bibfield  {journal} {\bibinfo  {journal}
  {Nature}\ }\textbf {\bibinfo {volume} {439}},\ \bibinfo {pages} {303}
  (\bibinfo {year} {2006})}\BibitemShut {NoStop}%
\bibitem [{\citenamefont {Ramirez}\ \emph {et~al.}(1999)\citenamefont
  {Ramirez}, \citenamefont {Hayashi}, \citenamefont {Cava}, \citenamefont
  {Siddharthan},\ and\ \citenamefont {Shastry}}]{ramirez1999zero}%
  \BibitemOpen
  \bibfield  {author} {\bibinfo {author} {\bibfnamefont {A.~P.}\ \bibnamefont
  {Ramirez}}, \bibinfo {author} {\bibfnamefont {A.}~\bibnamefont {Hayashi}},
  \bibinfo {author} {\bibfnamefont {R.~a.}\ \bibnamefont {Cava}}, \bibinfo
  {author} {\bibfnamefont {R.}~\bibnamefont {Siddharthan}}, \ and\ \bibinfo
  {author} {\bibfnamefont {B.}~\bibnamefont {Shastry}},\ }\href@noop {}
  {\bibfield  {journal} {\bibinfo  {journal} {Nature}\ }\textbf {\bibinfo
  {volume} {399}},\ \bibinfo {pages} {333} (\bibinfo {year}
  {1999})}\BibitemShut {NoStop}%
\bibitem [{\citenamefont {Castelnovo}\ \emph {et~al.}(2008)\citenamefont
  {Castelnovo}, \citenamefont {Moessner},\ and\ \citenamefont
  {Sondhi}}]{castelnovo2008magnetic}%
  \BibitemOpen
  \bibfield  {author} {\bibinfo {author} {\bibfnamefont {C.}~\bibnamefont
  {Castelnovo}}, \bibinfo {author} {\bibfnamefont {R.}~\bibnamefont
  {Moessner}}, \ and\ \bibinfo {author} {\bibfnamefont {S.~L.}\ \bibnamefont
  {Sondhi}},\ }\href@noop {} {\bibfield  {journal} {\bibinfo  {journal}
  {Nature}\ }\textbf {\bibinfo {volume} {451}},\ \bibinfo {pages} {42}
  (\bibinfo {year} {2008})}\BibitemShut {NoStop}%
\bibitem [{\citenamefont {M{\'o}l}\ \emph {et~al.}(2009)\citenamefont
  {M{\'o}l}, \citenamefont {Silva}, \citenamefont {Silva}, \citenamefont
  {Pereira}, \citenamefont {Moura-Melo},\ and\ \citenamefont
  {Costa}}]{mol2009magnetic}%
  \BibitemOpen
  \bibfield  {author} {\bibinfo {author} {\bibfnamefont {L.}~\bibnamefont
  {M{\'o}l}}, \bibinfo {author} {\bibfnamefont {R.}~\bibnamefont {Silva}},
  \bibinfo {author} {\bibfnamefont {R.}~\bibnamefont {Silva}}, \bibinfo
  {author} {\bibfnamefont {A.}~\bibnamefont {Pereira}}, \bibinfo {author}
  {\bibfnamefont {W.}~\bibnamefont {Moura-Melo}}, \ and\ \bibinfo {author}
  {\bibfnamefont {B.}~\bibnamefont {Costa}},\ }\href@noop {} {\bibfield
  {journal} {\bibinfo  {journal} {Journal of Applied Physics}\ }\textbf
  {\bibinfo {volume} {106}},\ \bibinfo {pages} {063913} (\bibinfo {year}
  {2009})}\BibitemShut {NoStop}%
\bibitem [{\citenamefont {Ladak}\ \emph {et~al.}(2010)\citenamefont {Ladak},
  \citenamefont {Read}, \citenamefont {Perkins}, \citenamefont {Cohen},\ and\
  \citenamefont {Branford}}]{ladak2010direct}%
  \BibitemOpen
  \bibfield  {author} {\bibinfo {author} {\bibfnamefont {S.}~\bibnamefont
  {Ladak}}, \bibinfo {author} {\bibfnamefont {D.}~\bibnamefont {Read}},
  \bibinfo {author} {\bibfnamefont {G.}~\bibnamefont {Perkins}}, \bibinfo
  {author} {\bibfnamefont {L.}~\bibnamefont {Cohen}}, \ and\ \bibinfo {author}
  {\bibfnamefont {W.}~\bibnamefont {Branford}},\ }\href@noop {} {\bibfield
  {journal} {\bibinfo  {journal} {Nature Physics}\ }\textbf {\bibinfo {volume}
  {6}},\ \bibinfo {pages} {359} (\bibinfo {year} {2010})}\BibitemShut {NoStop}%
\bibitem [{\citenamefont {Loreto}\ \emph {et~al.}(2015)\citenamefont {Loreto},
  \citenamefont {Morais}, \citenamefont {de~Araujo}, \citenamefont
  {Moura-Melo}, \citenamefont {Pereira}, \citenamefont {Silva}, \citenamefont
  {Nascimento},\ and\ \citenamefont {M{\'o}l}}]{loreto2015emergence}%
  \BibitemOpen
  \bibfield  {author} {\bibinfo {author} {\bibfnamefont {R.~P.}\ \bibnamefont
  {Loreto}}, \bibinfo {author} {\bibfnamefont {L.}~\bibnamefont {Morais}},
  \bibinfo {author} {\bibfnamefont {C.}~\bibnamefont {de~Araujo}}, \bibinfo
  {author} {\bibfnamefont {W.}~\bibnamefont {Moura-Melo}}, \bibinfo {author}
  {\bibfnamefont {A.}~\bibnamefont {Pereira}}, \bibinfo {author} {\bibfnamefont
  {R.}~\bibnamefont {Silva}}, \bibinfo {author} {\bibfnamefont
  {F.}~\bibnamefont {Nascimento}}, \ and\ \bibinfo {author} {\bibfnamefont
  {L.}~\bibnamefont {M{\'o}l}},\ }\href@noop {} {\bibfield  {journal} {\bibinfo
   {journal} {Nanotechnology}\ }\textbf {\bibinfo {volume} {26}},\ \bibinfo
  {pages} {295303} (\bibinfo {year} {2015})}\BibitemShut {NoStop}%
\bibitem [{\citenamefont {Chern}\ \emph {et~al.}(2014)\citenamefont {Chern},
  \citenamefont {Reichhardt},\ and\ \citenamefont
  {Nisoli}}]{chern2014realizing}%
  \BibitemOpen
  \bibfield  {author} {\bibinfo {author} {\bibfnamefont {G.-W.}\ \bibnamefont
  {Chern}}, \bibinfo {author} {\bibfnamefont {C.}~\bibnamefont {Reichhardt}}, \
  and\ \bibinfo {author} {\bibfnamefont {C.}~\bibnamefont {Nisoli}},\
  }\href@noop {} {\bibfield  {journal} {\bibinfo  {journal} {Applied Physics
  Letters}\ }\textbf {\bibinfo {volume} {104}},\ \bibinfo {pages} {013101}
  (\bibinfo {year} {2014})}\BibitemShut {NoStop}%
\bibitem [{\citenamefont {Perrin}\ \emph {et~al.}(2016)\citenamefont {Perrin},
  \citenamefont {Canals},\ and\ \citenamefont
  {Rougemaille}}]{perrin2016extensive}%
  \BibitemOpen
  \bibfield  {author} {\bibinfo {author} {\bibfnamefont {Y.}~\bibnamefont
  {Perrin}}, \bibinfo {author} {\bibfnamefont {B.}~\bibnamefont {Canals}}, \
  and\ \bibinfo {author} {\bibfnamefont {N.}~\bibnamefont {Rougemaille}},\
  }\href@noop {} {\bibfield  {journal} {\bibinfo  {journal} {Nature}\ }\textbf
  {\bibinfo {volume} {540}},\ \bibinfo {pages} {410} (\bibinfo {year}
  {2016})}\BibitemShut {NoStop}%
\bibitem [{\citenamefont {Nascimento}\ \emph {et~al.}(2012)\citenamefont
  {Nascimento}, \citenamefont {M{\'o}l}, \citenamefont {Moura-Melo},\ and\
  \citenamefont {Pereira}}]{nascimento2012confinement}%
  \BibitemOpen
  \bibfield  {author} {\bibinfo {author} {\bibfnamefont {F.}~\bibnamefont
  {Nascimento}}, \bibinfo {author} {\bibfnamefont {L.}~\bibnamefont {M{\'o}l}},
  \bibinfo {author} {\bibfnamefont {W.}~\bibnamefont {Moura-Melo}}, \ and\
  \bibinfo {author} {\bibfnamefont {A.}~\bibnamefont {Pereira}},\ }\href@noop
  {} {\bibfield  {journal} {\bibinfo  {journal} {New Journal of Physics}\
  }\textbf {\bibinfo {volume} {14}},\ \bibinfo {pages} {115019} (\bibinfo
  {year} {2012})}\BibitemShut {NoStop}%
\bibitem [{\citenamefont {Ribeiro}\ \emph {et~al.}(2017)\citenamefont
  {Ribeiro}, \citenamefont {Nascimento}, \citenamefont {Ferreira},
  \citenamefont {Moura-Melo}, \citenamefont {Costa}, \citenamefont {Borme},
  \citenamefont {Freitas}, \citenamefont {Wysin}, \citenamefont {Araujo},\ and\
  \citenamefont {Pereira}}]{ribeiro2017realization}%
  \BibitemOpen
  \bibfield  {author} {\bibinfo {author} {\bibfnamefont {I.}~\bibnamefont
  {Ribeiro}}, \bibinfo {author} {\bibfnamefont {F.}~\bibnamefont {Nascimento}},
  \bibinfo {author} {\bibfnamefont {S.}~\bibnamefont {Ferreira}}, \bibinfo
  {author} {\bibfnamefont {W.}~\bibnamefont {Moura-Melo}}, \bibinfo {author}
  {\bibfnamefont {C.}~\bibnamefont {Costa}}, \bibinfo {author} {\bibfnamefont
  {J.}~\bibnamefont {Borme}}, \bibinfo {author} {\bibfnamefont
  {P.}~\bibnamefont {Freitas}}, \bibinfo {author} {\bibfnamefont
  {G.}~\bibnamefont {Wysin}}, \bibinfo {author} {\bibfnamefont
  {C.}~\bibnamefont {Araujo}}, \ and\ \bibinfo {author} {\bibfnamefont
  {A.}~\bibnamefont {Pereira}},\ }\href@noop {} {\bibfield  {journal} {\bibinfo
   {journal} {Scientific reports}\ }\textbf {\bibinfo {volume} {7}},\ \bibinfo
  {pages} {13982} (\bibinfo {year} {2017})}\BibitemShut {NoStop}%
\bibitem [{\citenamefont {Bramwell}(2012)}]{bramwell2012magnetic}%
  \BibitemOpen
  \bibfield  {author} {\bibinfo {author} {\bibfnamefont {S.~T.}\ \bibnamefont
  {Bramwell}},\ }\href@noop {} {\bibfield  {journal} {\bibinfo  {journal}
  {Nature Physics}\ }\textbf {\bibinfo {volume} {8}},\ \bibinfo {pages} {703}
  (\bibinfo {year} {2012})}\BibitemShut {NoStop}%
\bibitem [{\citenamefont {Blundell}(2012)}]{blundell2012monopoles}%
  \BibitemOpen
  \bibfield  {author} {\bibinfo {author} {\bibfnamefont {S.~J.}\ \bibnamefont
  {Blundell}},\ }\href@noop {} {\bibfield  {journal} {\bibinfo  {journal}
  {Physical review letters}\ }\textbf {\bibinfo {volume} {108}},\ \bibinfo
  {pages} {147601} (\bibinfo {year} {2012})}\BibitemShut {NoStop}%
\bibitem [{\citenamefont {Porod}\ \emph {et~al.}(2014)\citenamefont {Porod},
  \citenamefont {Bernstein}, \citenamefont {Csaba}, \citenamefont {Hu},
  \citenamefont {Nahas}, \citenamefont {Niemier},\ and\ \citenamefont
  {Orlov}}]{porod2014nanomagnet}%
  \BibitemOpen
  \bibfield  {author} {\bibinfo {author} {\bibfnamefont {W.}~\bibnamefont
  {Porod}}, \bibinfo {author} {\bibfnamefont {G.~H.}\ \bibnamefont
  {Bernstein}}, \bibinfo {author} {\bibfnamefont {G.}~\bibnamefont {Csaba}},
  \bibinfo {author} {\bibfnamefont {S.~X.}\ \bibnamefont {Hu}}, \bibinfo
  {author} {\bibfnamefont {J.}~\bibnamefont {Nahas}}, \bibinfo {author}
  {\bibfnamefont {M.~T.}\ \bibnamefont {Niemier}}, \ and\ \bibinfo {author}
  {\bibfnamefont {A.}~\bibnamefont {Orlov}},\ }in\ \href@noop {} {\emph
  {\bibinfo {booktitle} {Field-Coupled Nanocomputing}}}\ (\bibinfo  {publisher}
  {Springer},\ \bibinfo {year} {2014})\ pp.\ \bibinfo {pages}
  {21--32}\BibitemShut {NoStop}%
\bibitem [{\citenamefont {Prinz}(1998)}]{prinz1998magnetoelectronics}%
  \BibitemOpen
  \bibfield  {author} {\bibinfo {author} {\bibfnamefont {G.~A.}\ \bibnamefont
  {Prinz}},\ }\href@noop {} {\bibfield  {journal} {\bibinfo  {journal}
  {Science}\ }\textbf {\bibinfo {volume} {282}},\ \bibinfo {pages} {1660}
  (\bibinfo {year} {1998})}\BibitemShut {NoStop}%
\bibitem [{\citenamefont {Engel}\ \emph {et~al.}(2005)\citenamefont {Engel},
  \citenamefont {Akerman}, \citenamefont {Butcher}, \citenamefont {Dave},
  \citenamefont {DeHerrera}, \citenamefont {Durlam}, \citenamefont
  {Grynkewich}, \citenamefont {Janesky}, \citenamefont {Pietambaram},
  \citenamefont {Rizzo} \emph {et~al.}}]{engel20054}%
  \BibitemOpen
  \bibfield  {author} {\bibinfo {author} {\bibfnamefont {B.}~\bibnamefont
  {Engel}}, \bibinfo {author} {\bibfnamefont {J.}~\bibnamefont {Akerman}},
  \bibinfo {author} {\bibfnamefont {B.}~\bibnamefont {Butcher}}, \bibinfo
  {author} {\bibfnamefont {R.}~\bibnamefont {Dave}}, \bibinfo {author}
  {\bibfnamefont {M.}~\bibnamefont {DeHerrera}}, \bibinfo {author}
  {\bibfnamefont {M.}~\bibnamefont {Durlam}}, \bibinfo {author} {\bibfnamefont
  {G.}~\bibnamefont {Grynkewich}}, \bibinfo {author} {\bibfnamefont
  {J.}~\bibnamefont {Janesky}}, \bibinfo {author} {\bibfnamefont
  {S.}~\bibnamefont {Pietambaram}}, \bibinfo {author} {\bibfnamefont
  {N.}~\bibnamefont {Rizzo}},  \emph {et~al.},\ }\href@noop {} {\bibfield
  {journal} {\bibinfo  {journal} {IEEE Transactions on Magnetics}\ }\textbf
  {\bibinfo {volume} {41}},\ \bibinfo {pages} {132} (\bibinfo {year}
  {2005})}\BibitemShut {NoStop}%
\bibitem [{\citenamefont {Rodrigues}\ and\ \citenamefont
  {M{\'o}l}(2018)}]{rodrigues2018towards}%
  \BibitemOpen
  \bibfield  {author} {\bibinfo {author} {\bibfnamefont {J.}~\bibnamefont
  {Rodrigues}}\ and\ \bibinfo {author} {\bibfnamefont {L.}~\bibnamefont
  {M{\'o}l}},\ }\href@noop {} {\bibfield  {journal} {\bibinfo  {journal}
  {Journal of Magnetism and Magnetic Materials}\ }\textbf {\bibinfo {volume}
  {458}},\ \bibinfo {pages} {327} (\bibinfo {year} {2018})}\BibitemShut
  {NoStop}%
\end{thebibliography}%

\end{document}